\documentclass[aps,prb,twocolumn,superscriptaddress,showpacs,amsmath,amssymb]{revtex4}
\usepackage{amsfonts}
\usepackage{graphicx}
\usepackage{dcolumn}
\usepackage{bm}
\usepackage{color}

\def\DR{\rm I\kern-1.45pt\rm R}
\def\DC{\kern2pt {\hbox{\sqi I}}\kern-4.2pt\rm C}

\newcommand{\ba}{\begin{array}}
\newcommand{\ea}{\end{array}}
\newcommand{\be}{\begin{equation}}
\newcommand{\ee}{\end{equation}}
\newcommand{\bea}{\begin{eqnarray}}
\newcommand{\eea}{\end{eqnarray}}
\newcommand{\bi}{\begin{itemize}}
\newcommand{\ei}{\end{itemize}}

\newcommand{\ra}{\rangle}
\newcommand{\la}{\langle}

\begin{document}

\title{Lattice distortions in a sawtooth chain with Heisenberg and Ising bonds.}
\author{ Stefano Bellucci}
\affiliation{INFN-Laboratori Nazionali di Frascati, Via E. Fermi 40, 00044 Frascati, Italy}
\author{Vadim Ohanyan}
\affiliation{Yerevan State University, A.Manoogian, 1, Yerevan,
0025 Armenia\\
 Yerevan Physics Institute, Alikhanian Br.2,
Yerevan, 0036, Armenia}



\begin{abstract}
An exactly solvable model of the sawtooth chain with Ising and
Heisenberg bonds and with coupling to lattice distortion for
Heisenberg bonds is considered in the magnetic field. Using the
direct transfer-matrix formalism an exact description of the
thermodynamic functions is obtained. The ground state phase diagrams
for all regions of parameters values containing phases corresponding
to the magnetization plateaus at $M=0,1/4$ and $1/2$ have been
obtained. Exact formulas for bond distortions for various ground
states are presented. A mechanism of magnetization plateau
stabilization with doubling of unit cell is reported. This mechanism
is related with the inequivalence of right and left interaction
bonds for certain lattice sites.
\end{abstract}

\date{\today}


\maketitle


\section{Introduction.}

The role of lattice distortions in the behavior of magnetic systems continues to be in the focus of intensive theoretical and
experimental investigations during last decade. The lift of the ground state degeneracy in frustrated magnets and magnetization
plateaus stabilization  mechanisms have been investigated intensively in various lattice spin systems with spin--lattice
coupling\cite{penc04,vekua06,gazza07,cabra06,bissola07,mila}. The concept of adiabatic phonos\cite{kittel} yielding the effective spin
Hamiltonian with additional biquadratic interaction is one of the main approximations using to gain inside into the properties of strongly
 correlated spin systems, interacting with lattice vibrations. Thus, considering only magnetic properties of the system with spin-lattice
 interaction, one deals with the Heisenberg model with additional biquadratic terms with coupling constant depending on the parameters of
 spin-lattice coupling and spring constant of the bond. Also, biquadratic terms can arise in the spin Hamiltonian from the quadrupole-quadrupole
 interactions. The spin--lattice coupling in quasi-one-dimensional frustrated spin chain, or zigzag ladder, has been shown to generate a
  series of magnetization plateaus at various rational values of magnetization in addition to one at $M=1/3$ existing in pure spin
  system \cite{vekua06}. It was also demonstrated than spin-lattice coupling in zigzag ladder gives rise to a novel type on magnetic
  excitations carrying fractional spin\cite{gazza07}. In two--dimensional both classical and quantum spin models with spin--lattice
  coupling the enhancement of magnetization plateau stabilization as well as appearance of new ordered phases due to biquadratic interaction
  have been also reported for the $J_1-J_2-J_3$-model\cite{bissola07} as well as for Shastry--Sutherland lattice\cite{mila} and pyrochlore
  antiferromagnet\cite{penc04}.

\begin{figure}[h]
\includegraphics[width=\columnwidth]{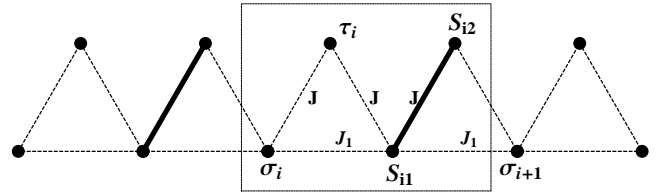}
\caption{The sawtooth chain. Heisenberg bonds are marked by thick lines, dashed lines represent Ising bonds.
Group of sites pictured into the frame correspond to one block.
\label{fig1}}
\end{figure}

In this paper we consider an exactly solvable one--dimensional spin  model with
Ising and Heisenberg bonds and spin--lattice interaction between the spins connected
 with Heisenberg bond. The geometry of the model corresponds to the system known as sawtooth chain.
 Sawtooth chain, or delta-chain, is one of the prototype examples of highly frustrated lattices.
 It has a structure of corner-sharing triangles (Fig.(\ref{fig1})). Unlike the other frustrated system,
 sawtooth chain has peculiar properties such us exactly known dimerized ground state and elementary
 excitations of quantum soliton type\cite{kub93, nak96, sen96, nak97, blu03}. Significance of the Heisenberg
 model on sawtooth chain is not limited only to academical interest. Magnetic lattices of few class of materials,
 such as delafossite YCuO$_{2.5}$\cite{del1,del2} and olivines with structure ZnL$_2$S$_4$(L=Er,Tm,Yb)\cite{oli} have
 been found to be of sawtooth chain type. In addition to that, recently much attention has been paid to the problem of
 localized magnon states or dispersionless excitation bands, which have been found in various frustrated spin and electron systems,
  particularly, in the Heisenberg and Hubbard models on sawtooth chain\cite{hon09,rich04,der07a,der07b,rich08,rich04a,schul02}.
  Calculating the possible ground state degeneracy corresponding to the localized magnon (or electron in case of Hubbard model)
  states one can describe low-temperature thermodynamics near saturation field for the corresponding frustrated system\cite{der07a,rich08}.
  However, exact description of thermodynamic properties for sawtooth chain as well as for many other strongly correlated lattice model
   is still an open issue. It is worth mentioning two sophisticated methods which allow one to construct thermodynamic functions of
   integrable model -- thermodynamic Bethe ansatz (TBA)\cite{tak} and quantum transfer-matrix method (QTM)\cite{klu}, however,
   applicability of these methods is limited to a very narrow class of integrable systems, such as $XXZ$-Heisenberg chain with
   homogeneous couplings, Hubbard chain, e.t.c.
\begin{figure}[tb]
\includegraphics[width=\columnwidth]{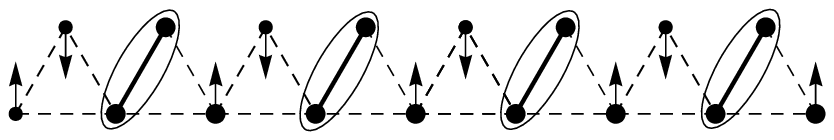}
\includegraphics[width=\columnwidth]{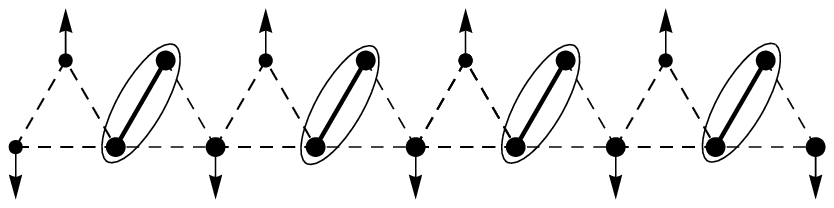}
\includegraphics[width=\columnwidth]{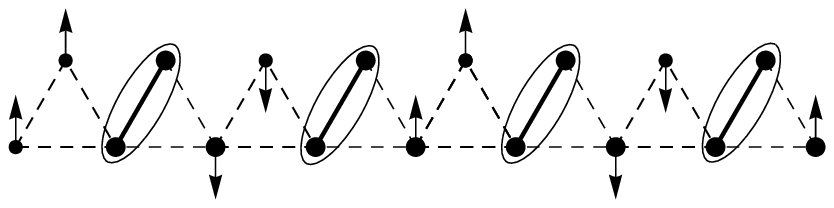}
\includegraphics[width=\columnwidth]{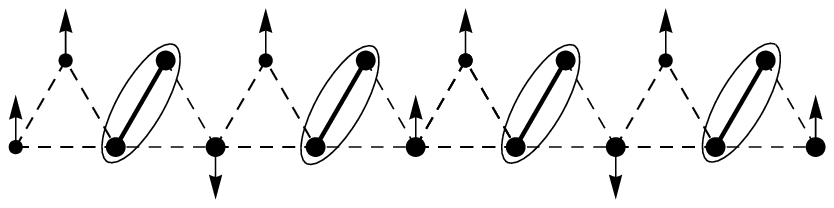}
\includegraphics[width=\columnwidth]{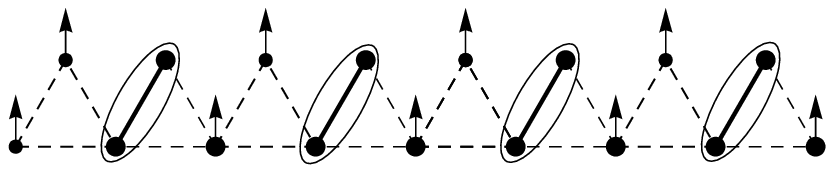}
\caption{Local spin configurations corresponding to (from up to bottom)
$|AF2\ra$, $|AF3\ra$, $|AF4\ra$, $|SM\ra$ and $|F1\ra$ ground states.
Ovals denote $|S^z_{tot}=0,+\ra$ singlet state of the pair of spins connected by $XXZ$ bond.}
\label{fig2ae}
\end{figure}
Despite great successes in exact describing of thermodynamic functions
for integrable models by TBA and especially by QTM\cite{klu2, trip}, for
many other physically and principally important low--dimensional strongly correlated
lattice models only laborious numerical calculations provide more or less reliable results
for finite $T$ thermodynamics. Recently, many papers have been devoted to exact solution of
the low--dimensional lattice spin models with mixed Ising and Heisenberg bonds\cite{str1, str2, str3, str4, str5, roj, ant09, oha08, oha09, per08, per09}
 or just to pure Ising counterparts of known Heisenberg models with various one--dimensional topologies of bonds\cite{oha03, lit, oha05, ayd1, ayd2}.
 These exact solutions for modified models have much in common with the numerical and experimental results obtained for their Heisenberg counterparts.
 For instance,  an alternating spin-chain with ferromagnetic--ferromagnetic--antiferromagnetic(FFA) interactions has qualitatively the same
 magnetization curve with magnetization plateau at $M=1/3$ for Heisenberg\cite{hida} and Ising spins\cite{oha03}.
 For more complicated F-F-AF-AF chain it has been demonstrated that replacing all ferromagnetic bonds with Ising ones,
 holding the rest (antiferromagnetic) bonds in original Heisenberg form does not lead to sufficient quantitative changes in
 the magnetization curve\cite{str1}. The corresponding mixed Ising--Heisenberg F-F-AF-AF chain has been proposed
 in Ref. [\onlinecite{str1}] as the model of magnetic structure of the compound Cu(3-Clpy)$_2$(N$_3$)$_2$,
 where 3-Clpy indicates 3-Chloropyridine. These two examples are the simplest cases of various one--dimensional
 spin lattice models with clusters of spins interacting with Heisenberg interaction and interacting to each other
 only through Ising bonds, which have been investigated intensively during last
 decade\cite{str1, str2, str3, str4, str5, roj, ant09, oha08, oha09, per08, per09}.
 Partition function and, thus, all thermodynamic functions for such  models can be obtained analytically.
 All these facts allow one to consider these models as an approximation for the underlying quantum spin systems.
 In Ref. [\onlinecite{oha09}] the corresponding approximate model has been considered for the sawtooth chain. Sawtooth
 chain with Ising and Heisenberg bonds considered there has four $S=1/2$ spins into the unit cell, thus, according to
 Oshikawa--Yamanaka--Affleck (OYA) criterion \cite{oya} it can exhibit magnetization plateaus at $M=0,1/4$ and $1/2$ what has been found
 in Ref.[\onlinecite{oha09}]. Plateau at $M=1/4$ appears due to difference into left and right interacting bonds for certain sites what,
 it its turn, results in unit cell doubling in the ground state corresponding to $M=1/4$.  In this paper we perform further investigations of this model complementing it with spin-lattice
  coupling and giving detailed analysis of all regions of the coupling constants values. The paper is organized as follows: in the Second section
   the Hamiltonian formulation of the model and its exact solution with classical transfer matrix methods is presented, the zero
   temperature phase diagrams are obtained in the third section, the fourth section contains an analysis of the average site
   displacements for different phases, concluding remarks are presented in the last section.

\section{Hamiltonian and thermodynamic functions}

We start with considering the sawtooth chain with $XXZ$ Heisenberg bond for the left pair of spins on every second triangle, while all other bonds are taken to be of Ising type. For the sake of briefness, hereafter, we will refer to these bonds as to quantum ones. Beside, each quantum bond is assumed to include elastic energy and displacement depended exchange constant (See Fig. \ref{fig1}). Under these assumptions the Hamiltonian can be decomposed into the sum of block Hamiltonians, commuting to each other:
\begin{widetext}
\begin{eqnarray}
&&\mathcal{H}=\sum_{i=1}^N\left(\mathcal{H}_i+J \sigma_i \tau_i-H\left(\tau_i+\frac{1}{2}\left( \sigma_i +\sigma_{i+1}\right) \right) \right), \nonumber \\
&&\mathcal{H}_i=J \left(1-A \rho_i \right)\left(\Delta\left(S_{i1}^xS_{i2}^x+S_{i1}^yS_{i2}^y \right)+S_{i1}^zS_{i2}^z \right)-
\left(H-J_1\left( \sigma_i+\sigma_{i+1}\right)-J\tau_i \right)S_{i1}^z-(H-J\sigma_{i+1})S_{i2}^z+\frac{K \rho_i^2}{2}, \label{ham}
\end{eqnarray}
\end{widetext}
where $N$ is the number of two-triangle blocks in the chain, $J_1$ and $J$ are coupling constants in the basement and between the side sites and basement respectively, $\rho_i$  is the distortion of quantum bond on $i$-th block, $A$ is dimensionless spin--lattice coupling, $K$ is the spring constant and $H$ is magnetic field assumed to point in $z$ direction. $S_{i(1,2)}^{\alpha}$ stand for $\alpha$ component of $S=1/2$ first(second) spin operator of $i-$th block and arrangement of Ising $\sigma$ and $\tau$ spins is depicted in Fig. (\ref{fig1}).
Due to commutating block structure of the Hamiltonian the partition function of the model can be represented in a way, suitable for applying the transfer-matrix formalism. Namely, one can expand the exponential and obtain the product of the terms, with each one corresponding to one block:
\begin{widetext}
\begin{eqnarray}
\mathcal{Z}=\sum_{\left(\sigma, \tau \right)}\mbox{Sp}_{\mathbf{S}}\prod_{i=1}^N \int_{-\infty}^{\infty} \frac{d \rho_i}{2 \pi} \exp \left(-\beta \left(\mathcal{H}_i+J \sigma_i \tau_i-H\left(\tau_i+\frac{1}{2}\left( \sigma_i +\sigma_{i+1}\right) \right) \right) \right). \label{Z}
\end{eqnarray}
\end{widetext}
 Here the sum is over all values of Ising variables and $\mbox{Sp}_{\mathbf{S}}$ stands for the trace over all states of quantum spins, for each quantum bond there is also integral over all possible values of distortion. After implementing integration, one arrives at the effective Hamiltonian with additional biquadratic term, which is widely used in the investigations of spin systems with adiabatic coupling to phonons \cite{penc04,vekua06,gazza07,cabra06,bissola07,mila,kittel}. Then, trace over quantum states can be taken independently for each block, yielding
 \begin{widetext}
 \begin{eqnarray}
 \mathcal{Z}=\sum_{\left(\sigma, \tau \right)}\prod_{i=1}^N\frac{1}{\sqrt{2 \pi K \beta}}
 \mbox{Sp}\large[\exp\left( -\beta\mathcal{H}_i^{eff}\right)\large]\exp\left(-\beta\left(J \sigma_i \tau_i-H\left(\tau_i+\frac{1}{2}\left( \sigma_i +\sigma_{i+1}\right) \right) \right)\right),  \label{Z2}
 \end{eqnarray}
 \end{widetext}
 where effective Hamiltonian reads
 \begin{widetext}
 \begin{eqnarray}
 &&\mathcal{H}_i^{eff}=J \left(\mathbf{S}_{i1}\mathbf{S}_{i2} \right)_{\Delta}-b\left(\mathbf{S}_{i1}\mathbf{S}_{i2} \right)^2_{\Delta}-\left(H-J_1\left( \sigma_i+\sigma_{i+1}\right)-J\tau_i \right)S_{i1}^z-(H-J\sigma_{i+1})S_{i2}^z, \nonumber \\
 &&\left(\mathbf{S}_{i1}\mathbf{S}_{i2} \right)_{\Delta}=\Delta\left(S_{i1}^xS_{i2}^x+S_{i1}^yS_{i2}^y \right)+S_{i1}^zS_{i2}^z, \label{Heff}
 \end{eqnarray}
 \end{widetext}
 where the effective constant of biquadratic interaction is given by the relation $b=\frac{J^2A^2}{2K}$. As is known, biquadratic term in case of spin $s=1/2$ yields the constant and bilinear term. So, one should also have in mind the following form of the effective Hamiltonian:
 \begin{widetext}
 \bea
 &&\mathcal{H}_i^{eff}=-\frac{b}{16}\left(1+2 \Delta^2 \right)+J_{XY}\left(S_{i1}^xS_{i2}^x+S_{i1}^yS_{i2}^y \right)+J_{Z}S_{i1}^zS_{i2}^z-\left(H-J_1\left( \sigma_i+\sigma_{i+1}\right)-J\tau_i \right)S_{i1}^z-(H-J\sigma_{i+1})S_{i2}^z, \nonumber \\
 &&J_{XY}=\Delta\left( J+\frac{1}{2}b\right), \quad J_Z=J+\frac{b \Delta^2}{2}. \label{Heff2}
 \eea
 \end{widetext}
 Thus, one actually obtains the Hamiltonian of the $XXZ$-model with peculiar parametrization of the exchange constants and anisotropy. Any crucial changes in the properties of the system like appearance of novel phases which do not appear at $b=0$ does not expected, while spin--lattice coupling can lead to some non-trivial impact on the ground states phase diagrams properties.
 Then, the partition function of that type can be calculated exactly \cite{ant09, oha08, oha09} within transfer-matrix formalism. To proceed, one needs to calculate the trace for each block quantum spins, and then to sum over two values of $\tau_i$ in each block to give Eq. (\ref{Z2}) the standard form corresponding to the transfer-matrix technique\cite{bax}
\begin{widetext}
\begin{eqnarray}
&&\mathcal{Z}=\left(\frac{1}{\sqrt{2 \pi K \beta}} \right)^N\sum_{\left(\sigma\right)}\prod_{i=1}^N T\left( \sigma_i, \sigma_{i+1}\right)=\left(\frac{1}{\sqrt{2 \pi K \beta}} \right)^N\mbox{Sp}\mathbf{T}^N, \nonumber \\
&&T\left(\sigma_i, \sigma_{i+1}\right)=Z\left(\sigma_i, \sigma_{i+1} \right)e^{\beta\frac{H}{2}\left( \sigma_i, \sigma_{i+1}\right)}, \nonumber \\
&&Z\left(\sigma_i, \sigma_{i+1} \right)=\sum_{\tau_i=\pm1/2}\Omega\left(\sigma_i, \sigma_{i+1}|\tau_i\right)e^{-\beta \left( J\sigma_i-H\right)\tau_i}, \label{Z3}
\end{eqnarray}
\end{widetext}
where
\begin{widetext}
\bea
&&\Omega\left(\sigma_i, \sigma_{i+1}|\tau_i\right)=\sum_{n=1}^4 e^{-\beta \lambda_n\left(\sigma_i, \sigma_{i+1}|\tau_i \right)}=2e^{\beta\frac{b}{16}}\left(e^{-\beta \frac{J}{4}}\cosh\left(\beta \frac{1}{2}\left(2H-J\left(\tau_i+\sigma_{i+1}-J_1\left( \sigma_i+\sigma_{i+1}\right) \right) \right) \right)\right. \nonumber \\
&&\left.+e^{\beta \frac{1}{4}\left(\Delta^2 b +J \right)}\cosh\left( \beta \frac{1}{2}\sqrt{\left(J\left( \sigma_{i+1}-\tau_i  \right)-J_1 \left(\sigma_i + \sigma_{i+1} \right) \right)^2+
\Delta^2 \left( J +\frac{1}{2}b\right)^2}\right) \right),\label{Omega}
\end{eqnarray}
\end{widetext}
where $\lambda_n\left(\sigma_i, \sigma_{i+1}|\tau_i \right)$ are four eigenvalues of $\mathcal{H}_i^{eff}$:
\begin{widetext}
\begin{eqnarray}
&&\lambda_{1,2}\left(\sigma_i, \sigma_{i+1}|\tau_i \right)=\frac{1}{16}\left(4J-b \right)\label{lambda}
\pm\frac{1}{2}\left( J\left( \tau_i+\sigma_{i+1}\right)+J_1\left( \sigma_i+\sigma_{i+1}\right)\right)\mp H, \nonumber \\
&&\lambda_{3,4}\left(\sigma_i, \sigma_{i+1}|\tau_i \right)=-\frac{1}{16}\left(b\left( 1+4 \Delta^2\right)+4 J \right)\mp
\frac{1}{2}\sqrt{\left(J\left(\sigma_{i+1}-\tau_i  \right)-J_1 \left(\sigma_i + \sigma_{i+1} \right) \right)^2+
\Delta^2 \left( J +\frac{1}{2}b\right)^2}, \nonumber \\
\end{eqnarray}
\end{widetext}
The corresponding eigenstates are two polarized vectors with total spin projection $S^z_{tot}=\pm 1$ for $\lambda_{1,2}$ and two states with $S^z_{tot}=0$ for $\lambda_{3,4}$:
\begin{widetext}
\begin{eqnarray}
&&|S^z_{tot}=0, \pm \rangle=\frac{1}{\sqrt{1+\gamma_{\pm}^2}}\left(|\uparrow\downarrow\rangle-\gamma_{\pm}|\downarrow\uparrow\rangle \right), \nonumber \\
&& \gamma_{\pm}=\frac{J\left(\sigma_{i+1}-\tau_i  \right)-J_1 \left(\sigma_i + \sigma_{i+1} \right)
\pm  \sqrt{\left(J\left( \sigma_{i+1}-\tau_i  \right)-J_1 \left(\sigma_i + \sigma_{i+1} \right) \right)^2+
\Delta^2 \left( J +\frac{1}{2}b\right)^2}}{\Delta \left(J+\frac{1}{2}b\right)} . \label{gamma}
\end{eqnarray}
\end{widetext}
Thus, for calculating the partition function of the model, Eq. (\ref{Z3}), one should find eigenvalues of the transfer matrix $\mathrm{T}$,
\bea
{\mathbf{T}}=\left( \begin{array}{lcr}
      e^{\beta  \frac{H}{2}}Z_+  & Z_0 \\
      \widetilde{Z}_0  & e^{-\beta  \frac{H}{2}}Z_- \label{T}
      \end{array}
\right), \eea
where entries are connected with the value of block partition function as follows $Z_+=Z\left(1/2,1/2 \right), Z_0=Z\left(1/2, -1/2 \right), \widetilde{Z}_0=Z\left(-1/2,1/2 \right), Z_-=Z\left(-1/2,-1/2 \right)$. Having all that one can easily write the expression for the free energy per block in the thermodynamic limit, when only the maximal eigenvalue survives
\begin{widetext}
\bea
f=-\frac{1}{2\beta}\log\left(\frac{1}{2 \pi K \beta} \right)-\frac{1}{\beta}\log\frac{1}{2}\left(e^{\beta \frac{H}{2}}Z_+ + e^{-\beta \frac{H}{2}}Z_-
 +\sqrt{\left(e^{\beta \frac{H}{2}}Z_+ - e^{-\beta \frac{H}{2}}Z_- \right)^2+4 Z_0 \widetilde{Z}_0}\right). \label{f}
\eea
\end{widetext}
One can see that the first additive term accounts for the vibration contribution to free energy , while effects of interactions between spin degrees of freedom and lattice displacements are incorporated into the $b$-dependance of the rest part of free energy. After that, all thermodynamic functions can be found exactly by taking corresponding derivatives of the free energy.
For instance, for magnetization one obtains:
\bea
M=-\frac{1}{2}\left( \frac{\partial f}{\partial H}\right)_{T}. \label{M}
\eea

\section{Ground states and $T=0$ phase diagrams}

Analyzing possible magnetic ground states of the model under consideration at $T=0$ one finds large variety of spin configurations with a spacial period equal to the period of the chain. However, there is also another two ground states where translational symmetry is broken up to the period of two blocks. All states can be also classified by the values of magnetization, possible values of which are $0,1/2$ and $1$ for states with unbroken translational symmetry, and $0$ and $1/4$ for the state with doubling of lattice period. Accordingly, the magnetization curves with magnetization plateaus at all aforementioned values of magnetization are expected.
Let us start with antiferromagnetic spin configurations with $M=0$. There are three antiferromagnetic spin configurations with spacial period equal to lattice period in the system. In the first one all quantum spins are polarized, while both of their adjacent $\sigma$ and $\tau$ spins point in the opposite direction:

\bea
&&|AF1\rangle=\prod_{i=1}^N |\uparrow\uparrow\rangle_i\bigotimes|\sigma_i=\downarrow,\tau_i=\downarrow\rangle, \nonumber \\
&&\varepsilon_{AF1}=-\frac{1}{16}b-\frac{1}{2}J_1, \label{AF1}
\eea
here $\varepsilon$ is the corresponding energy per one block. Another two antiferromagnetic ground states of the system are degenerated by energy. They correspond to $|S_{tot}^z=0,+\rangle$ eigenstate (Eq. (\ref{gamma}) ) for each $XXZ$-bond and opposite orientation for two Ising spins in each block:
\begin{widetext}
\bea
&&|AF2\rangle=\prod_{i=1}^N|S_{tot}^z=0,+\rangle_i\bigotimes|\sigma_i=\uparrow,\tau_i=\downarrow\rangle, \nonumber \\
&&|AF3\rangle=\prod_{i=1}^N|S_{tot}^z=0,+\rangle_i\bigotimes|\sigma_i=\downarrow,\tau_i=\uparrow\rangle, \nonumber\\
&&\varepsilon_{AF(2,3)}=-\frac{1}{2}J-\frac{1}{16}b \left(1+4\Delta^2 \right)-\frac{1}{2}\sqrt{\left(J-J_1 \right)^2+\Delta^2\left(J+\frac{1}{2}b \right)^2}, \label{AM23}
\eea
\end{widetext}
where the value of the coefficient $\gamma$ in the ground state $|S_{tot}^z=0,+\rangle_i$ is different according to Eq. (\ref{gamma}).
There is also another antiferromagnetic ground state with broken translational symmetry. In this ground state pairs of quantum spins are in $|S_{tot}^z=0,+\rangle$ state while for each block $\sigma_i$=$\tau_i$ but their joint direction alternates from block to block:
\begin{widetext}
\bea
&&|AF4\rangle=\prod_{i=1}^{N/2}|S_{tot}^z=0,+\rangle_{2i-1}\bigotimes|\sigma_{2i-1}=\uparrow,\tau_{2i-1}=\uparrow\rangle\bigotimes
|S_{tot}^z=0,+\rangle_{2i}\bigotimes|\sigma_{2i}=\downarrow,\tau_{2i}=\downarrow\rangle, \nonumber \\
&&\varepsilon_{AF4}=-\frac{1}{16}b\left(1+4 \Delta^2 \right)-\frac{1}{2}\sqrt{J^2+\Delta^2\left(J+\frac{1}{2}b \right)^2}. \label{AF4}
\eea
\end{widetext}

The system demonstrates also three ferrimagnetic phases with value of magnetization equal to $1/2$:
\begin{widetext}
\bea
&&|F1\rangle=\prod_{i=1}^N|S_{tot}^z=0,+\rangle_i\bigotimes|\sigma_i=\uparrow,\tau_i=\uparrow\rangle, \nonumber \\
&& \varepsilon_{F1}=-\frac{1}{16} b \left( 1+4 \Delta^2\right)-\frac{1}{2}\sqrt{J_1^2+\Delta^2\left(J+\frac{1}{2}b \right)^2}-H, \nonumber \\
&&|F2\rangle=\prod_{i=1}^N|\uparrow\uparrow\rangle_i\bigotimes|\sigma_i=\downarrow,\tau_i=\uparrow\rangle, \nonumber\\
&&\varepsilon_{F2}=-\frac{1}{16}b - \frac{1}{2}J_1-H, \nonumber \\
&&|F3\rangle=\prod_{i=1}^N|\uparrow\uparrow\rangle_i\bigotimes|\sigma_i=\uparrow,\tau_i=\downarrow\rangle, \nonumber\\
&&\varepsilon_{F3}=-\frac{1}{16}b + \frac{1}{2}J_1-H. \label{FFF}
\eea
\end{widetext}
Local spin configuration corresponding to several non trivial ground sates which appear as the zero-temperature ground states of the system under consideration are presented in Fig. (\ref{fig2ae}).
Finally, let us mention the spin polarized, saturated state, where all spins are pointed along the external magnetic field
\bea
&&|SP\rangle=\prod_{i=1}^N |\uparrow\uparrow\rangle_i\bigotimes|\sigma_i=\uparrow,\tau_i=\uparrow\rangle, \nonumber \\
&&\varepsilon_{SP}=J+\frac{1}{2}J_1-\frac{1}{16} b -2H. \label{SP}
\eea
Within some region of the values of system parameters, which is specified below, there is also another state with broken translational symmetry. Namely, this is an intermediate state between $|AF2\rangle$ or $|AF3\rangle$ and $|F1\rangle$ where the $\sigma$ spins are ordered antiferromagnetically with respect to each other, while all $\tau$ spins are pointed along the field. This state can be achieved from $|AF2\rangle$ or $|AF3\rangle$ by flipping every second basement $\sigma$ spin. We will refer to this state as a spin-modulated stare as here the spacial modulation of local spin polarization is occurred along the chain. Namely, local spin polarization of each left pair of spins on each triangle varies in the following way $100010001000.....$.  The corresponding ground state and energy per one block are
\begin{widetext}
\bea
&&|SM\rangle=\prod_{i=1}^{N/2}|S_{tot}^z=0,+\rangle_{2i-1}\bigotimes|\sigma_{2i-1}=\uparrow,\tau_{2i-1}=\uparrow\rangle\bigotimes
|S_{tot}^z=0,+\rangle_{2i}\bigotimes|\sigma_{2i}=\downarrow,\tau_{2i}=\uparrow\rangle, \nonumber \\
&&\varepsilon_{SM}=-\frac{1}{16}b \left(1+4 \Delta^2 \right)-\frac{1}{4}J-\frac{1}{4}|\Delta\left(J+\frac{1}{2}b\right)|-\frac{1}{4}\sqrt{J^2+\Delta^2\left(J+\frac{1}{2}b \right)^2}-\frac{1}{2}H. \label{SM}
\eea
\end{widetext}
Another one ground state can be appeared in the sawtooth chain with Ising and Heisenberg bonds without lattice distortions\cite{oha09}. When $b=0$ and $J=J_1$ the ground state of the system is a macroscopically twofold degenerated frustrated state
\bea
|FR\rangle=\prod_{i=1}^N|S_{tot}^z=0,+\rangle\bigotimes|\xi_i\rangle,
\eea
where $|\xi_i\rangle$ stands for either $|\sigma_i=\uparrow,\tau_i=\downarrow\rangle$ or $|\sigma_i=\downarrow,\tau_i=\uparrow\rangle$. Thus, each pair of $\sigma-\tau$ spin in each block is frustrated and can freely pass from one possible state to another.
\begin{figure}[tb]
\includegraphics[width=\columnwidth]{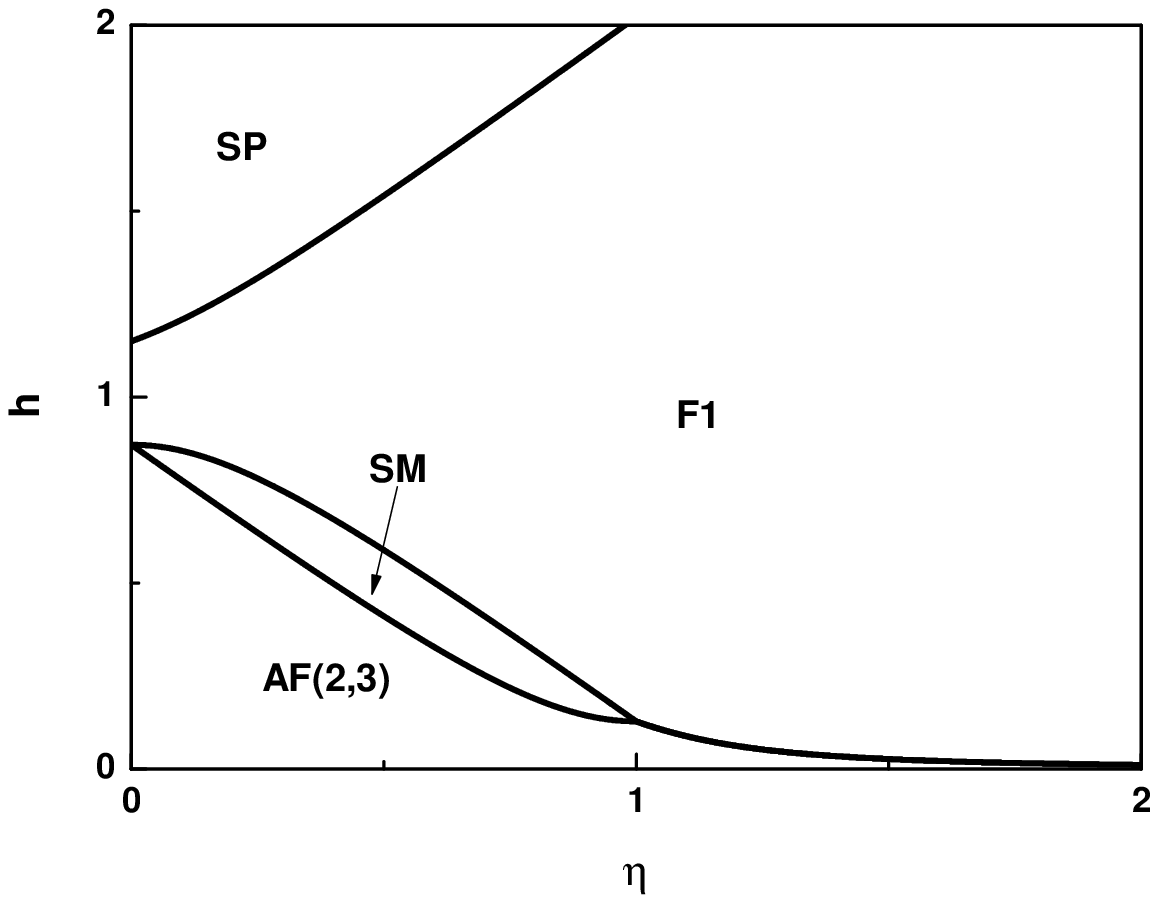}
\includegraphics[width=\columnwidth]{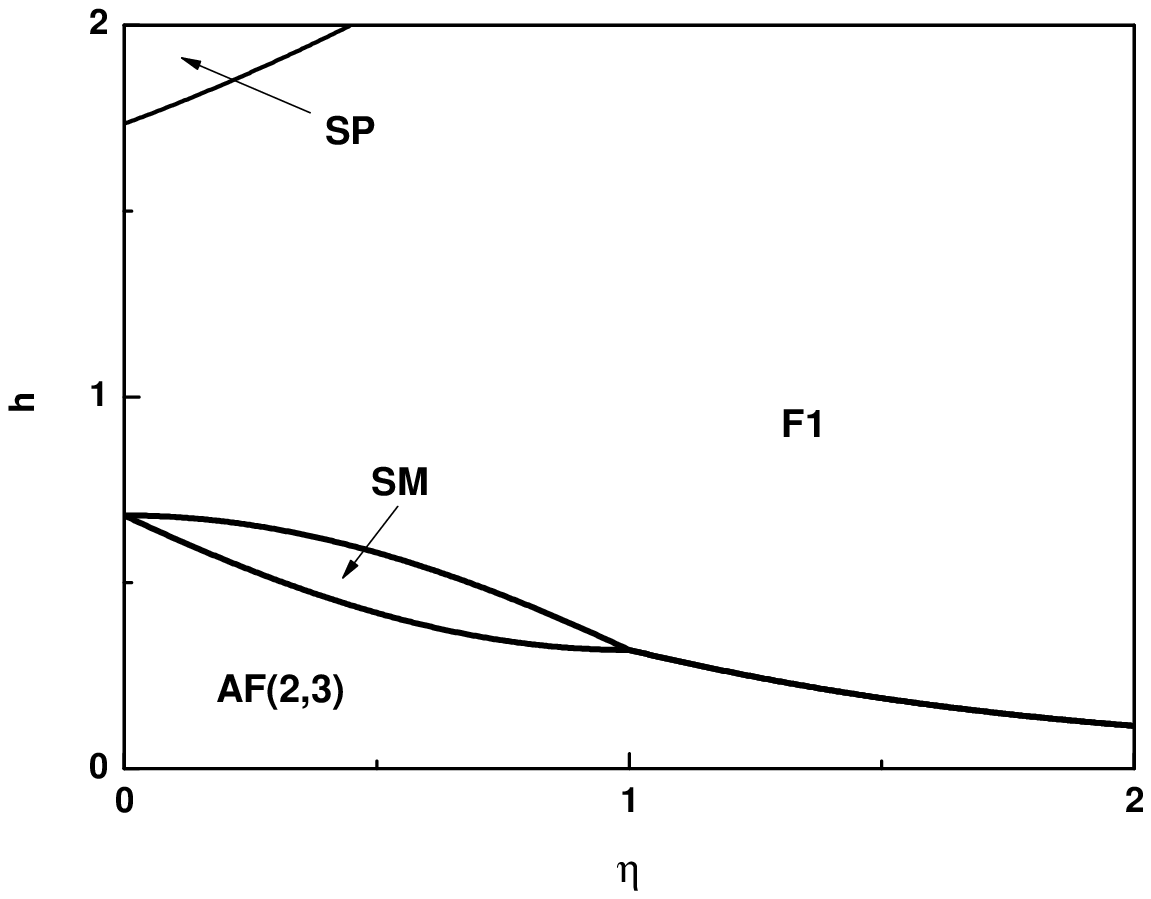}
\caption{Ground state $T=0$ phase diagrams in $(\eta=J_1/J,h=H/J)$-plane for antiferromagnetic coupling ($J>0$, $J_1>0$) for the sawtooth chain with Ising and Heisenberg bond without biquadratic term (upper panel); and for $b/J=6$ (lower panel). The corresponding eigenstates for the lattice are shown in Fig. (\ref{fig2})
\label{fig2}}
\end{figure}
\begin{figure}[tb]
\includegraphics[width=\columnwidth]{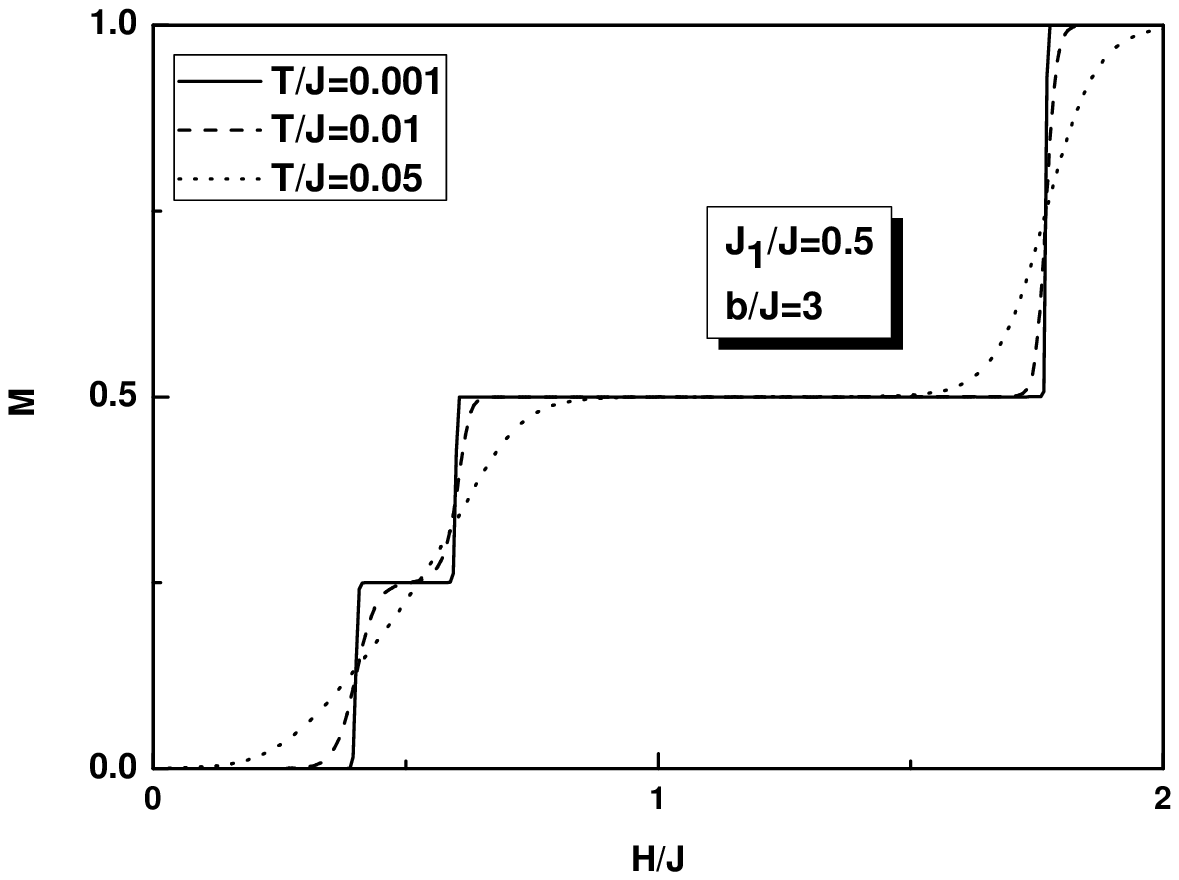}
\includegraphics[width=\columnwidth]{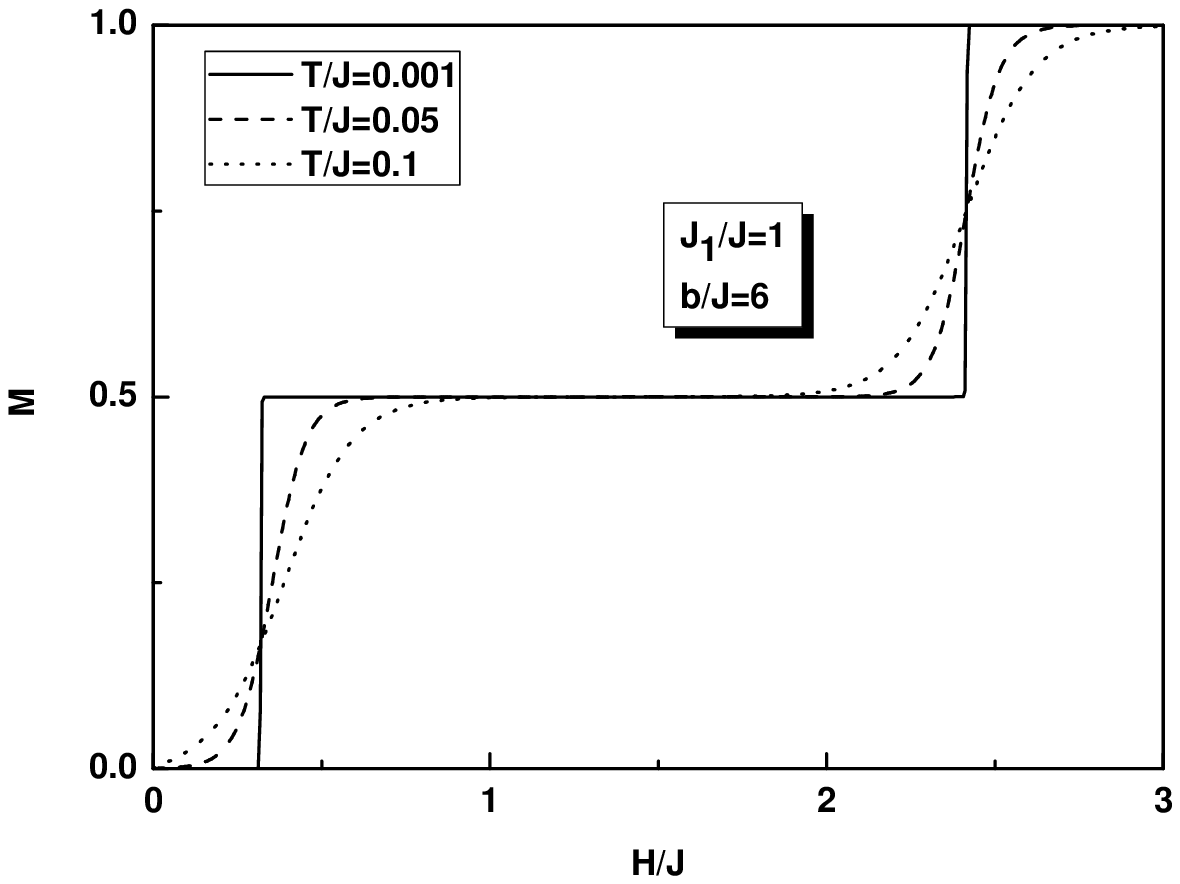}
\caption{The plots of magnetization processes for the antiferromagnetic region of interactions, ($J>0$, $J_1>0$), for $\eta=0.5$ (upper panel)  and
$\eta=1$ lower panel for several temperatures. One can see effect of thermal fluctuation on the spin-modulated phase, the plateau at $M=1/4$ shrink down very rapidly with increasing the temperature, while plateau at $M=1/2$ is much more stable.}
\label{fig3}
\end{figure}
\begin{figure}[tb]
\includegraphics[width=\columnwidth]{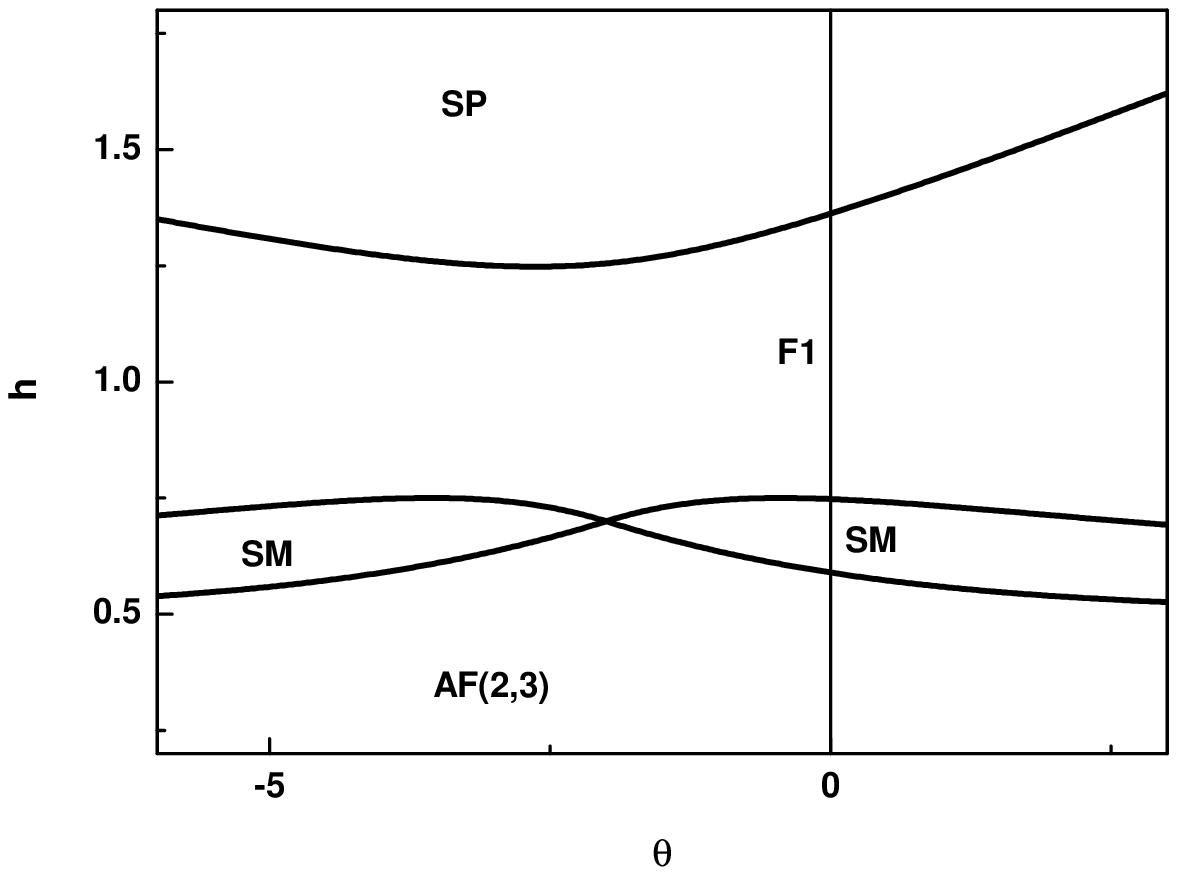}
\caption{The ground states phase diagram in $(\theta=b/J,h=H/J)$-plane demonstrating effect of biquadratic terms for $\eta=0.3$ and $\Delta=0.3$.}
\label{fig4}
\end{figure}

\subsection{Ground state phase diagram for $J>0$, $J_1>0$}

Let us first consider the case of purely antiferromagnetic couplings $J>0$, $J_1>0$. The value of the biquadratic interaction constant is assumed to be always positive, as $b=\frac{J^2A^2}{2 K}, K>0$.  At such values of parameters the system exhibits four different ground states, $|AF2\rangle$ or $|AF3\rangle$ which are degenerated by energies, $|F1\rangle$, $|SM\rangle$, which can appear only at $\eta=J_1/J<1$ and $|SP\rangle$. When the magnetic field in turned off the ground state of the system at $T=0$ is always an antiferromagnetic one, corresponding to either $|AF2\rangle$ or $|AF3\rangle$, which are closed analogs of dimerized ground states of conventional quantum sawtooth chain\cite{kub93, nak96, sen96}. However, in the exactly solvable model considered here the structure of ground state is more complicated. In contrast to the conventional case, instead of simple dimers, $\frac{1}{\sqrt{2}}\left( |\uparrow\downarrow\rangle-|\downarrow\uparrow\rangle\right)$, here one finds $|S_{tot}^z=0,+\rangle$ states alternating with pairs of oppositely directed Ising spins Eq.(\ref{AM23}). Then, with the magnitude of the magnetic field increasing, a sequence of quantum phase transitions takes place, which increase the magnetization of the system in stepwise manner. The corresponding phase diagrams in the $(\eta, h)$-plane are presented in Fig. (\ref{fig2}), the upper panel demonstrates phase diagram for the sawtooth chain with Ising and Heisenberg bond without lattice distortions, which was previously obtained in Ref. [~\onlinecite{oha09}], the lower panel demonstrates effects of biquadratic term originated by lattice distortions. Actually, this term enhances the magnetization plateau at $M=1/2$, which corresponds to the ferrimagnetic phase $|F1\rangle$. The region of phase diagram corresponding to the spin-modulated phase almost is not affected by the value of $\theta=b/J$. This can be explained by the fact, that spin-modulated phase which corresponds to the magnetization plateau at the value $M=1/4$, appears due to very special structure of the eigenstate of quantum spins with $S_{tot}^z=0$ which is strongly affected by the values of adjacent Ising spins. Absence of the permutational symmetry in the $|S_{tot}^z=0,+\rangle$ state and specific dependence of the corresponding eigenvalue on the values of adjacent Ising spins makes the state with alternating orientation of basement Ising spins stable for certain value of parameters, namely for $\eta<1$. Thus, this is an example of the novel mechanism of magnetization plateau formation which consists in the doubling of the spacial period of the ground state due to the difference in effective fields, $h_1$ and $h_2$ which act on $z$-components of two quantum spins in one block, namely $h_1=H-J\tau_i-J_1(\sigma_i+\sigma_{i+1})$, $h_2=H-J_1\sigma_{i+1}$. The equations of phase boundaries between four aforementioned phases are:

\begin{widetext}
\bea
&&\mbox{between} \quad |AF(2,3)\rangle \quad \mbox{and} \quad |SM\rangle :\quad h=\frac{1}{2}\left(1-|D_+\left( \theta \right)|-\sqrt{1+D_+^2\left( \theta \right)} \right)+\sqrt{\left(1-\eta\right)^2+D_+^2\left( \theta \right) },\label{b} \\
&&\mbox{between} \quad |AF(2,3)\rangle \quad \mbox{and} \quad |F1\rangle :\quad h=\frac{1}{2}\left(1+\sqrt{\left(1-\eta\right)^2+D_+^2\left( \theta \right)}-\sqrt{\eta^2+D_+^2\left( \theta \right)} \right), \nonumber \\
&&\mbox{between} \quad |SM\rangle \quad \mbox{and} \quad |F1\rangle :\quad h=\frac{1}{2}\left(1+|D_+\left( \theta \right)|+\sqrt{1+D_+^2\left( \theta \right)}\right)-\sqrt{\eta^2+D_+^2\left( \theta \right)} , \nonumber \\
&&\mbox{between} \quad |F1\rangle \quad \mbox{and} \quad |SP\rangle :\quad h=1+\frac{1}{2}\left(\eta+\sqrt{\eta^2+D_+^2\left( \theta \right)} \right)+\frac{1}{4}\Delta^2 \theta, \nonumber
\eea
where
\bea
D_+^2\left( \theta \right)=\Delta^2 \left(1+\frac{1}{2}\theta \right)^2. \label{D}
\eea
\end{widetext}
Two typical plots of magnetization processes corresponding to all antiferromagnetic  for different temperatures  coupling are presented in Fig. (\ref{fig3}), upper panel shows the $\eta<1$ with three magnetization plateaus at $M=0,1/4$ and $1/2$, the lower panel corresponds to the $\eta>1$ case with only two plateaus at $M=0$ and $M=1/2$. As one can see spin-modulated phase in much more sensitive to thermal fluctuations than the rest plateaus, because it is connected with the configuration of Ising spins which is less stable that the entangled state of quantum spins.

Looking at Fig. (\ref{fig2}) one can also notice the effect of biquadratic interaction on the length of plateaus. It almost does not affect the plateau at $M=1/4$ as spin-modulated phase appears due to spin-flip of Ising spins in the basement, however, minor deformations of the region corresponding to the spin-modulated phase can be observed from the phase diagrams. In its turn, biquadratic interaction enhances the plateau at $M=1/2$ corresponding to $|F1\ra$ phase as well as the plateau at $M=0$ corresponding to $|AF(2,3)\ra$ phase. The effects of biquadratic term or lattice distortions are summarized in Fig. (\ref{fig4}), where the phase diagram in the $(\theta,h)$-plane is presented. One can see almost parallel arrangement of the four phases mentioned above and one peculiar point at $\theta=-2$ where spin-modulated phase disappears. Negative values of $b$ are irrelevant in the model of lattice distortions, as $b=\frac{J^2A^2}{2 K}$ and $K>0$. However, formally one can consider negative values as well.  This effect is obviously connected with the peculiar properties of Hamiltonian with biquadratic interaction $\mathcal{H}=J_1 \left( \mathbf{S}_1 \mathbf{S}_2\right)+J_2 \left( \mathbf{S}_1 \mathbf{S}_2\right)^2$ at $J_2=2J_1$. At this values all four eigenvectors became separable, corresponding just to the standard basis $|\uparrow\uparrow\ra, |\downarrow\downarrow\ra, |\uparrow\downarrow\ra, |\downarrow\uparrow\ra$, which makes impossible realizations of spin-modulated phase.
\subsection{Ground state phase diagram for $J<0$, $J_1>0$}
\begin{figure}[tb]
\includegraphics[width=\columnwidth]{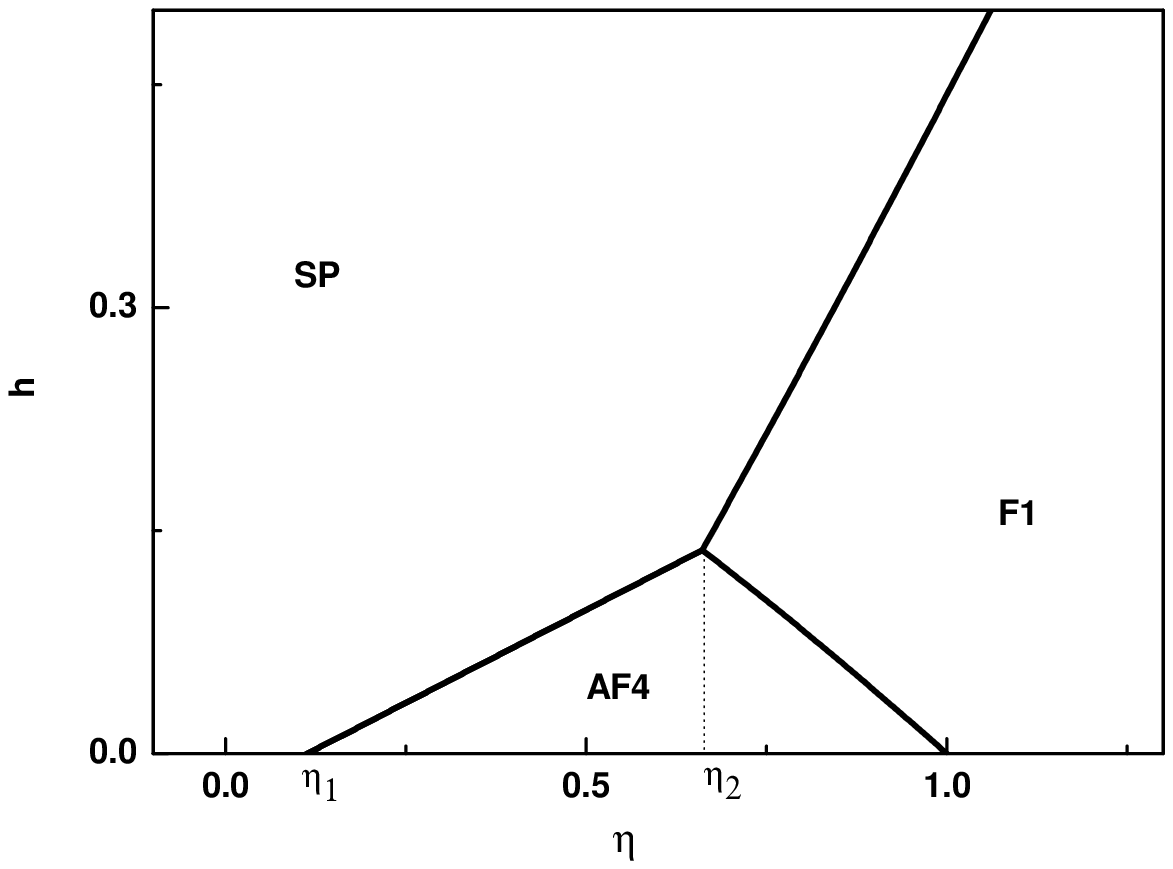}
\includegraphics[width=\columnwidth]{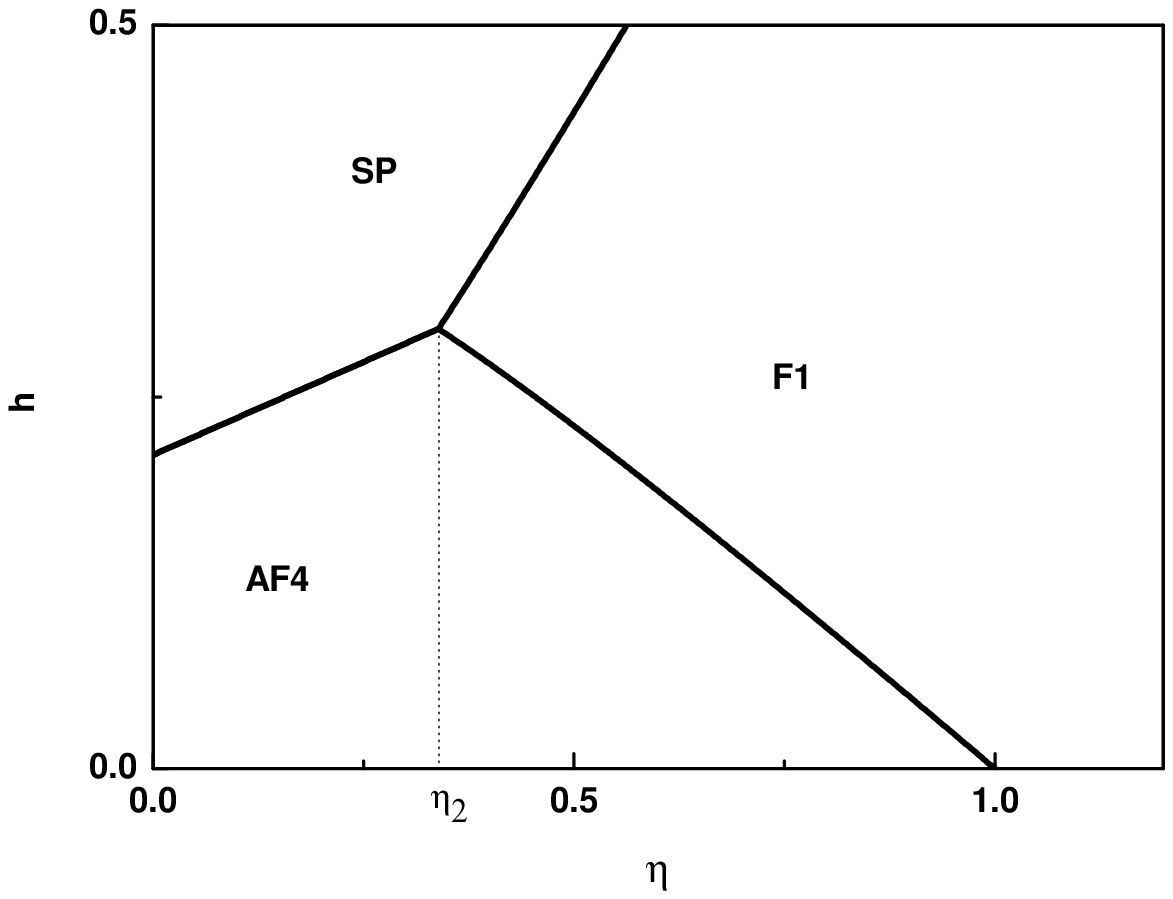}
\includegraphics[width=\columnwidth]{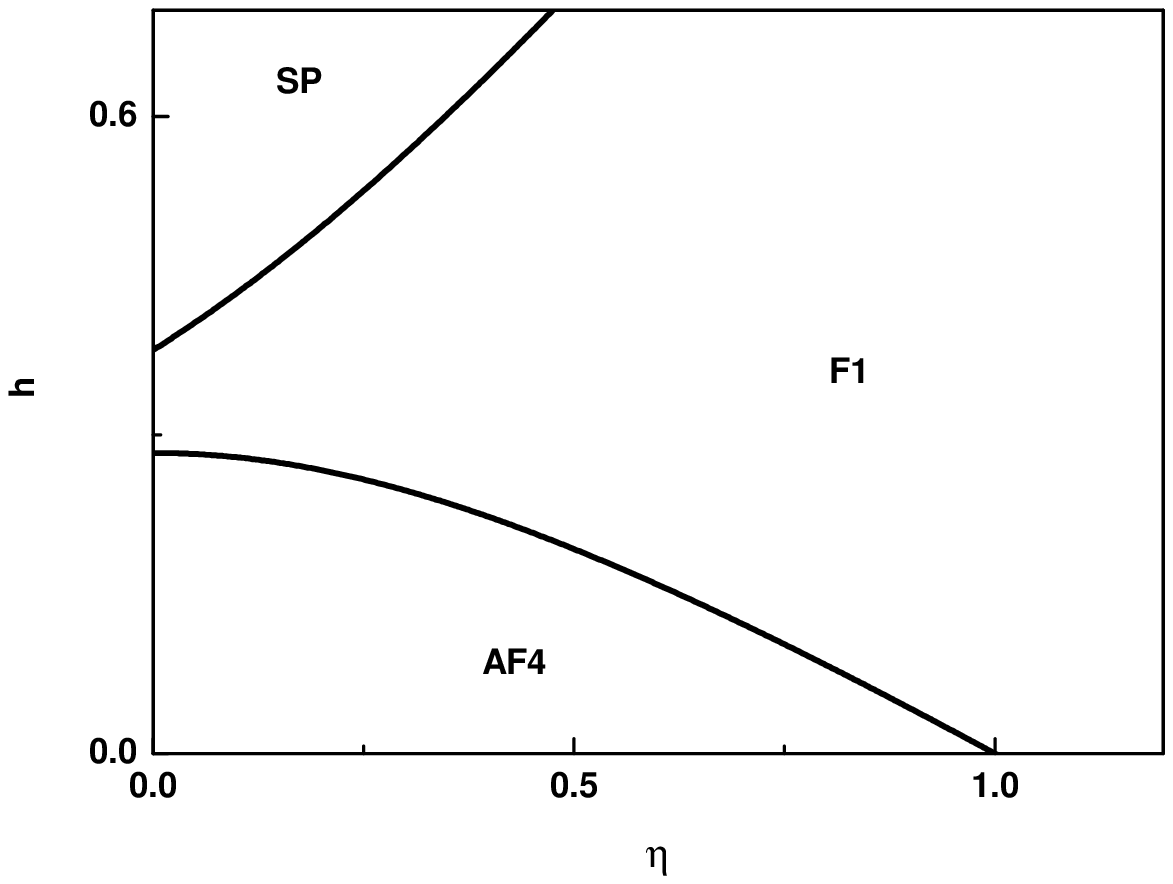}
\caption{The ground states phase diagram for $J=-1$, $\Delta=1.2$ and $\theta=1$(upper panel), $\theta=2.5$ (middle panel) and $\theta=3$ (lower panel). Here $h=H/|J|, \theta=b/|J|$.}
\label{fig5}
\end{figure}

Let us consider now another region of the coupling constants values corresponding to ferromagnetic $J$. In contrast to the previous case here three different ground states are possible at $H=0$ depending on the value of $\eta=\frac{J_1}{|J|}$. It is interesting to note, that the only one phase with zero magnetization (antiferromagnetic) which appears in this case is $|AF4\rangle$ configuration with broken translational symmetry. Depending of the values of $\theta=\frac{b}{|J|}$ and $\Delta$ the ground state phase diagram can have three qualitatively different forms (Fig \ref{fig5}), mainly determined by positions of two special points in $\eta$ axis. One of them, $\eta_1$, determines the quantum phase transition point between spin polarized and antiferromagnetic $|AF4\rangle$ phases in the absence of magnetic field. Another one, $\eta_2$ corresponds to the position of quantum triple point in the $(\eta, h)$-plane:
\begin{widetext}
\bea
&&\eta_1=2-\frac{\Delta^2 \theta}{2}-\sqrt{1+D_-^2\left( \theta \right)}, \nonumber \\
&&\eta_2=\frac{1}{3}\left( \frac{\Delta^2 \theta}{2}-2-\sqrt{1+D_-^2\left( \theta \right)}+2\sqrt{5+4\sqrt{1+D_-^2\left( \theta \right)}
-\Delta^2 \left(2+\theta \left(\frac{1}{2}\theta+\sqrt{1+D_-^2\left( \theta \right)} \right) \right)+\frac{\Delta^4 \theta^2}{4}}\right), \nonumber \\
&&D_-^2\left( \theta\right)=\Delta^2 \left( 1-\frac{1}{2}\theta\right)^2. \label{eta12}
\eea
\end{widetext}
There is also another special point in the $\eta$-axis corresponding to the quantum phase transition point between antiferromagnetic and ferrimagnetic ground states at $H=0$. This value is parameter independent and is equal to one. When $\eta>1$ the ground state of the system is $|F1\rangle$. One can see that at positive values of $\theta$ always $\eta_1<\eta_2<1$. Thus, when $\eta_1>0$ one obtains the ground state phase diagram with spin polarized phase for $\eta<\eta_1$, antiferromagnetic phase $|AF4\rangle$ turning immediately to spin polarized phase under the action of magnetic field for $\eta_1<\eta<\eta_2$, antiferromagnetic phase $|AF4\rangle$ turning to the ferrimagnetic phase $|F1\ra$ prior to spin polarized phase for $\eta_2<\eta<1$ and ferrimagnetic phase turning immediately to spin polarized phase for $\eta>1$. One can see the corresponding phase diagram for particular value $\theta=1$ in the upper panel of Fig. (\ref{fig5}). Once $\eta_1$ became negative, then one will get another topology of ground state phase diagram presented in middle panel of Fig. (\ref{fig5}) for $\theta=2.5$. And, finally, when $\eta_2$ becomes negative only two ground states for $H=0$ are possible, $|AF4\ra$ and $|F1\ra$ and there is no quantum triple point in the phase diagram (see lower panel of Fig. (\ref{fig5}), where $\theta=3$). The equations of phase boundaries between three phases occurring in the negative $J$ case are
\begin{widetext}
\bea
&&\mbox{between} \quad |AF4\rangle \quad \mbox{and} \quad |F1\rangle :\quad h=\frac{1}{2}\left(\sqrt{1+D_-^2\left( \theta \right)}-\sqrt{\eta^2+D_-^2\left( \theta \right)} \right),\label{b2} \\
&&\mbox{between} \quad |AF4\rangle \quad \mbox{and} \quad |SP\rangle :\quad h=-1+\frac{1}{2}\left(\eta+\frac{\Delta^2 \theta}{2}+\sqrt{1+D_-^2\left( \theta \right)} \right), \nonumber \\
&&\mbox{between} \quad |F1\rangle \quad \mbox{and} \quad |SP\rangle :\quad h=-1+\frac{1}{2}\left(\eta+\frac{\Delta^2 \theta}{2}+\sqrt{\eta^2+D_-^2\left( \theta \right)} \right). \nonumber \\
\eea
\end{widetext}

\subsection{Ground state phase diagram for $J>0$, $J_1<0$}
\begin{figure}[tb]
\includegraphics[width=\columnwidth]{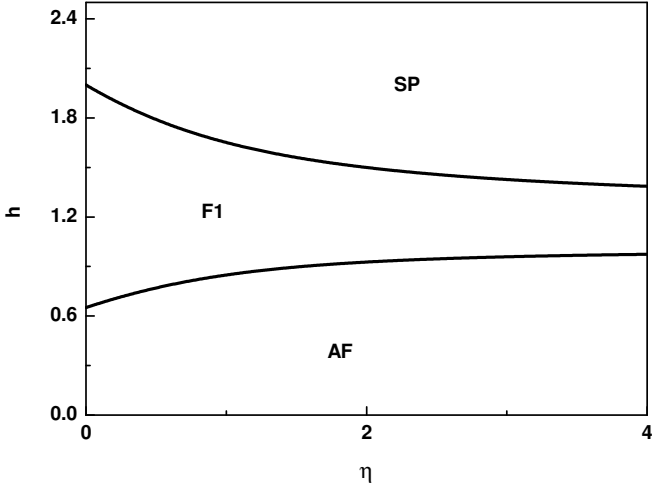}
\caption{The ground states phase diagram for $J_1=-1$, $J=1$, $\theta=1$ and $\Delta=1$. Here $\eta=|J_1|/J$, $h=H/J$.}
\label{fig6}
\end{figure}

At the ferromagnetic value of the coupling constant between the basement spins $J_1<0$, the system exhibits rather simple ground states phase diagram with almost parallel arrangement of three phases, $|AF(2,3)\ra$, $|F1\ra$ and $|SP\ra$. The $H=0$ ground state is always antiferromagnetic which with increase of the external magnetic filed magnitude turned to the ferrimagnetic phase $|F1\ra$ with further transition to spin polarized saturated phase. The corresponding phase diagram is presented on Fig. (\ref{fig6}). Accordingly, the phase boundaries are given by the same equation as of the purely antiferromagnetic case $J>0$, $J_1>0$ with replacement $\eta$ by $-\eta$. (See Eq.(\ref{b})). Varying parameter of biquadratic interaction $b$ does not lead to a crucial change in the general picture.
\section{Average displacement}
\begin{figure}[tb]
\includegraphics[width=\columnwidth]{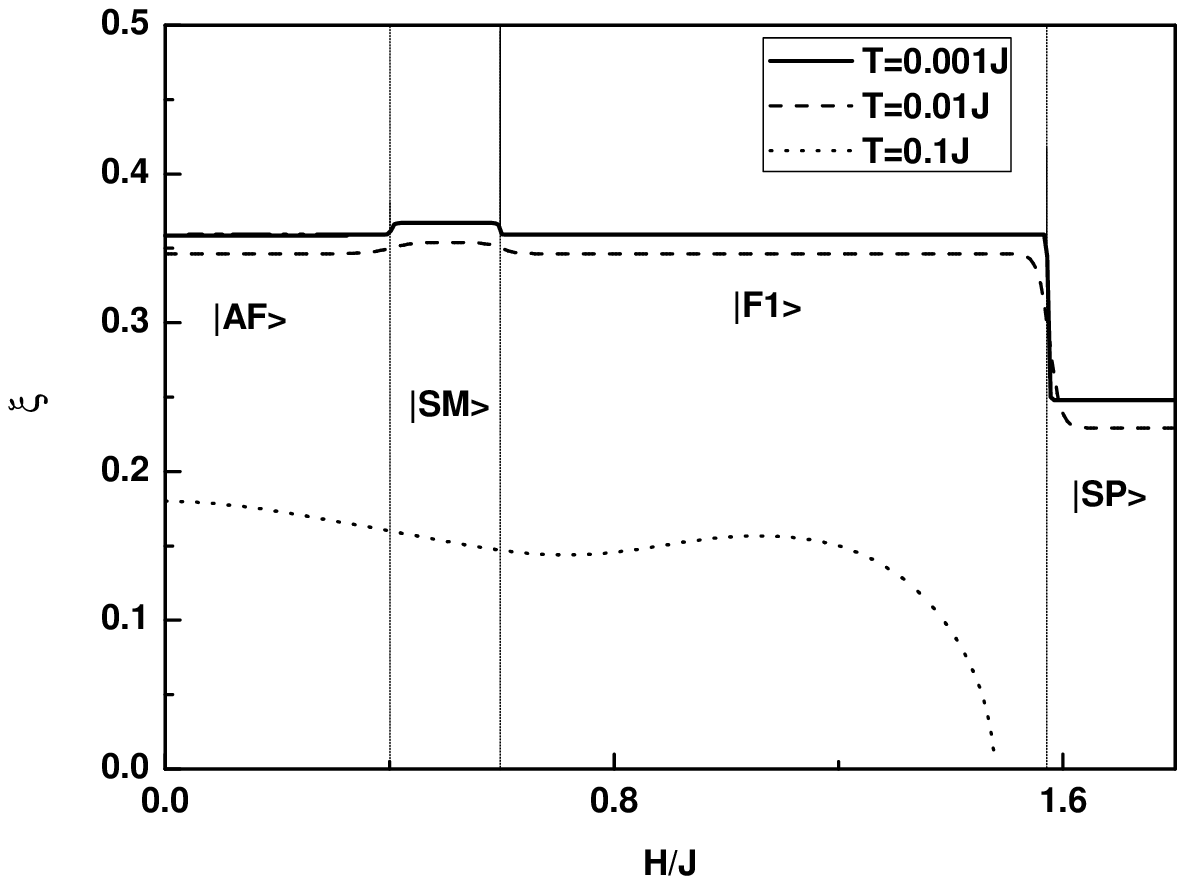}
\caption{Standard deviation of the distance between sites with quantum spins vs. external magnetic field magnitude for three different temperatures and $J=1$, $J_1=0.5$, $b=0.5$ and $\Delta=0.3$. Regions of different ground states corresponding to $T=0$ phase diagram are separated from each other by horizontal thin lines.}
\label{fig7}
\end{figure}

One can easily determine the expression for the equilibrium point of the quantum bond distortion immediately minimizing the block Hamiltonian (\ref{ham}):
\bea
\hat{\rho}_0=\frac{J A}{K}\left( \mathbf{S}_1 \mathbf{S}_2\right)_{\Delta}. \label{dis}
\eea
Thus, in the equilibrium, the distance between sites with quantum spins is determined by their magnetic state. In order to calculate the bond displacement for various ground states one just needs to take quantum-mechanical average of this operator. One can also exploit thermodynamical identities to determine standard  deviation of $\rho$, more precisely
\bea
\xi=\sqrt{\frac{\sum_{i=1}^N \rho_i^2}{N}}=\sqrt{2\left(\frac{\partial f}{\partial K}\right)_{T,H}}. \label{xi}
\eea
On the other hand, one can obtain zero temperature values of distance between the sites with quantum spins just by calculating quantum mechanical averages of the operator $\hat{\rho}_0=\frac{1}{N}\sum_{i=1}^N \hat{\rho}_{i0}$ for corresponding ground states of the system. For all ground states where pairs of quantum spins are in $|\uparrow\uparrow\ra$ state, i.e. for $|AF1\ra$, $|F2\ra$, $|F3\ra$ and $|SP\ra$ one obtains
\bea
\langle \hat{\rho}_0 \rangle=\frac{J A}{4 K}, \label{avrho1}
\eea
while for other ground states average distance between quantum spins is given by more complicated expressions depending of the orientation of the other spins of the block. For the ground states with unbroken block--translational symmetry one obtains
\bea
&&\rho_0=\la \Psi|{\hat{\rho}}_0 |\Psi \ra = 
\label{avrho2} \\
&&-\frac{J A}{4 K}\left(1+\frac{2 \Delta}{\sqrt{1+\left(\tilde{J}(\tau-\sigma_R)-\tilde{J_1}(\sigma_L+\sigma_R) \right)^2}} \right)\nonumber
\eea
where the following notations are adopted
\bea
\tilde{J}=\frac{J}{\Delta\left(J+\frac{1}{2}b\right)}, \quad \tilde{J_1}=\frac{J_1}{\Delta\left(J+\frac{1}{2}b\right)}, \label{til}
\eea
and $|\Psi \ra$ stands for the one of the $|AF(2,3)\ra$, $|F1\ra$ ground states, and $\gamma_{\Psi}$ is the coefficient from the $|S_{tot}^z=0,+\ra$ (Eq. (\ref{gamma})) state calculated for the corresponding values of $\sigma_{L(R)}$ and $\tau$, where $\sigma_{L(R)}$ is the value of the left(right) sigma-spin surrounding quantum spin pairs in the given ground state. For the ground states with block doubling, $|AF4\ra$ and $|SP\ra$ the corresponding expression has the form
\bea
\la \Psi|{\hat{\rho}}_0 |\Psi \ra=\frac{1}{2}\left(\rho_1+\rho_2 \right), \label{avrho3}
\eea
where $\rho_{1,2}$ are calculated for the first(second) block in the two-block unit cell of the ground states configuration. Each $\rho_i$ is given by Eq. (\ref{avrho2}) with corresponding values of $\sigma_{L(R)}$ and $\tau$. In order to compare zero-temperature results with thermodynamically obtained expression form Eq.  (\ref{xi}) one needs also to calculate quantum--mechanical averages for the square of operator $\hat{\rho}_0$. Then, one will obtain a quantity thermal average of which coincides with $\xi$ from Eq. (\ref{xi}). Defining $\xi_0=\sqrt{\la \Psi|{\hat{\rho^2}}_0|\Psi\ra}$ one will obtain for ground states without translational symmetry breaking
\bea
&&\xi_0= \label{xi0}\\
&&\frac{J A}{4 K}\sqrt{1+4 \Delta \left(\Delta+\frac{1}{\sqrt{1+\left(\tilde{J}(\tau-\sigma_R)-\tilde{J_1}(\sigma_L+\sigma_R) \right)^2}} \right)}, \nonumber
\eea
and for $|AF4\ra$ and $|SP\ra$ phases
\bea
\sqrt{\la \Psi|{\hat{\rho^2}}_0|\Psi\ra}=\sqrt{\frac{1}{2}\left(\xi_1^2+\xi_2^2 \right)}, \label{xi12}
\eea
where $\xi_{1(2)}$ are calculated according to Eq. (\ref{xi0}) for left and right triangles in the block. Below the list of distances between the site with quantum spins for different phases for zero temperature is presented
\begin{widetext}
\bea
\label{dis2}
&&|AF(2,3)\ra: \quad \rho_0=-\frac{JA}{4K}\left( 1+\frac{2 \Delta}{\sqrt{1+\left(\tilde{J}+\tilde{J_1} \right)^2}}\right), \quad
\xi_0=\frac{JA}{4K}\sqrt{1+4 \Delta\left(\frac{1}{\sqrt{1+\left(\tilde{J}+\tilde{J_1} \right)^2}} +\Delta\right)}, \label{rhoxi} \\
&&|AF4\ra: \quad \rho_0=-\frac{JA}{4K}\left( 1+\frac{2 \Delta}{\sqrt{1+\tilde{J}^2}}\right), \quad \xi_0=\frac{JA}{4K}\sqrt{1+4 \Delta\left(\frac{1}{\sqrt{1+\tilde{J}^2}} +\Delta\right)}, \nonumber \\
&&|F1\ra: \quad \rho_0=-\frac{JA}{4K}\left( 1+\frac{2 \Delta}{\sqrt{1+\tilde{J_1}^2}}\right), \quad
\xi_0=\frac{JA}{4K}\sqrt{1+4 \Delta\left(\frac{1}{\sqrt{1+\tilde{J_1}^2}} +\Delta\right)},\nonumber \\
&&|SM\ra: \quad \rho_1=-\frac{JA}{4K}\left( 1+\frac{2 \Delta}{\sqrt{1+\tilde{J}^2}}\right),\quad
\rho_2=-\frac{J A}{4 K}\left(1+2 \Delta \right),\quad \rho_0=-\frac{J A}{4 K}\left(1+\Delta \left( 1+\frac{1}{\sqrt{1+\tilde{J}^2}}\right) \right)\nonumber, \\
&& \xi_1=\frac{J A}{4 K}\sqrt{1+4\Delta \left( \frac{1}{\sqrt{1+\tilde{J}^2}}+\Delta\right)}, \quad \xi_2=\frac{J A}{4 K}\sqrt{ 1+4 \Delta (1+\Delta)}, \quad \xi_0=\frac{JA}{4K}\sqrt{1+2\Delta \left(\frac{1}{\sqrt{1+\tilde{J}^2}}+2 \Delta \right)}.\nonumber
\eea
\end{widetext}
Thus, from Eqs. (\ref{avrho1}) and (\ref{dis2}) one can conclude that for $J>0$ the interplay between elastic properties of the quantum bonds and magnetic behavior od spins connected by it yields the stretching of the bond length for those ground states in which $S_{tot}^z=1$ for each quantum bonds ($|AF1\rangle, |F2\rangle, |F3\rangle$ and $|SP\rangle$), while for the ground states  with $S_{tot}^z=0$ ($|AF(2,3)\rangle, |AF4\rangle, |F1\rangle$ and $|SM\rangle$) antiferromagnetic coupling of spins connected by $XXZ$ bond leads to shortening of bond length. Though, in later case the expressions for the average displacement are moro complicated and contain all interaction parameters of the system, only the sign of $J$ and $\Delta$ defines whether the equilibrium displacement will be positive or negative.
Magnetic field dependence of the standard deviation of the distance between lattice sites with quantum spins given by Eq. (\ref{xi}) is presented in Fig. (\ref{fig7}). One can see low temperature step-like changes of $\xi$ according to the connection between site distance and magnetic state of spin situated at them (Eq. (\ref{dis})).

\section{Concluding remarks}

In this paper we presented a complete analysis of the magnetic properties of the sawtooth chain with
 Ising and Heisenberg bond and spin--lattice coupling for spins interacting with Heisenberg interaction.
 After integration over site displacements one deals with the additional biquadratic spin interaction. Due to special arrangement
  of Ising and Heisenberg bonds exact calculation of partition function of the system has been performed.
  The system exhibits large variety of ordered phases, among which are those with the doubling of the unit cell.
  Magnetic behavior of the system is also rather rich. Depending of the values of parameters, coupling constant in the
  basement $J_1$, coupling constant for the interaction between spins on the top and basement spins $J$ and effective
  coupling constant of additional biquadratic interaction $b$, various magnetization curves are possible with
   magnetization plateaus at $M=0, 1/4$ and $1/2$. Let us mention, that exact diagonalization calculations for
   ordinary antiferromagnetic sawtooth chain at $J=2J_1$ reported in Ref. [\onlinecite{rich04a}], revealed only
   one plateau at $M=1/2$, which corresponds to the so-called magnon crystal, an eigenstate with whole filling of all
   possible localized magnon states. In our case, the microscopic physical origin of this plateau state is completely different
   due to special structure of the interactions. However, at corresponding values of coupling constants (the left half of the phase
   diagram presented in Fig. (\ref{fig2}) in upper panel) the system with mixed Ising and Heisenberg bonds considered here exhibits
    magnetization curve with all three magnetization plateaus mentioned above.
    These features
    have obvious origin. The ordinary sawtooth chain\cite{rich04a} has unit cell with
    two $S=1/2$ spins, thus, according to OYA criterion\cite{oya}, it can display only plateaus at $M=0$ and
    $M=1/2$, while the system considered in this paper has four
    $S=1/2$ spin in the unit cell which can result in additional magnetization
    plateau at $M=1/4$. However, OYA criterion\cite{oya} only specify
    possible values of magnetization plateaus for given type on
    translational symmetry and given values of spin. For stabilization of the possible plateaus additional physical mechanisms are required. In the system considered here a special attention should be paid to the magnetization
    plateau at $M=1/4$. The origin of the corresponding eigenstate $|SM\ra$ is in the nonequivalence of the left and right $\sigma$ spin
    for each pair of quantum spins, provided later are in the $|S_{tot}^z=0,+\ra$ state in which $\mathbf{S}_1$ and $\mathbf{S}_2$ in their
    turn are nonequivalent. Thus, at a certain value of external magnetic field magnitude, the eigenstate with doubling of the unit cell
    becomes stable, in which $\sigma$ spins in the left and right blocks of the doubled unit cell are in different states. To our knowledge,
    this is the novel mechanism of magnetization plateau stabilization inherent in the systems with mixed Ising and Heisenberg bonds.
    The same arguments are valid for the antiferromagnetic state with doubling unit cell $|AF4\ra$. Generally speaking, the same mechanism
    can be presented in the purely quantum models with inhomogeneous interaction. For sawtooth chain, one can consider the model
    where coupling constants in the basement, on the left leg and on the right leg of the triangles, are all different, or consider
    the model with alternating interaction in the basement. Probably, for some values of coupling constant the magnetic behavior of
    such a sawtooth chain will have much in common with that shown here for model with Ising and Heisenberg bonds, particularly, magnetization
    plateau at $M=1/4$. This issue requires further investigation in order to clarify the deep connections between magnetic and thermodynamic
    properties of quantum spin models and their Ising-Heisenberg counterparts. There is surprisingly good correspondence between magnetization
    curves of quantum F-F-AF-AF alternating chain and the same chain with ferromagnetic bonds changed with the Ising ones obtained
    in Ref. [\onlinecite{str1}]. This, as well as other
    results\cite{str1, str2, str3, str4, str5, roj, ant09, oha08, oha09, per08, per09,oha03, lit, oha05, ayd1, ayd2}, allows one
    to consider the change of interaction bonds with Ising ones for providing exact solvability, as an approximate methods in theory of
    strongly correlated spin lattice models. An especially good agreement can be achieved when only ferromagnetic bonds are changed with
    Ising ones, because the ground state of two ferromagnetically interacting spins is the same for Ising and for Heisenberg interaction.
    While, for antiferromagnetically interacting spins quantum ground state is spin singlet which is an entangled state which has no direct
    analogies in case of Ising interaction. Thus, changing antiferromagnetic bond with Ising one leads to crucial loose of important physical
    properties, while for ferromagnetic coupling the physical difference is not so pronounced.



\acknowledgments
We are grateful to J. Stre\v{c}ka for valuable comments.
V.O. expresses his gratitude to LNF-INFN for hospitality during the work on the paper and acknowledges partial support form the grants
  CRDF-UCEP - 06/07 and ANSEF-1518-PS.


\end{document}